\documentclass[twocolumn,showpacs,preprintnumbers,superscriptaddress,amsmath,amssymb]{revtex4}
\usepackage{graphicx}
\usepackage{dcolumn}
\usepackage{bm}
\usepackage{color}
\usepackage{booktabs, longtable}
\usepackage{multirow}
\usepackage{mathptmx}
\usepackage{CJK}
\usepackage{ltxtable}
\usepackage[colorlinks,linkcolor=blue,anchorcolor=blue,citecolor=red]{hyperref}

\begin{document}

\title{Continuum Skyrme-Hartree-Fock-Bogoliubov theory with Green's function method\\ for odd-$A$ nuclei}

\author{Ting-Ting Sun}
\email{ttsunphy@zzu.edu.cn}
\affiliation{School of Physics and Engineering, Zhengzhou University, Zhengzhou 450001, China}

\author{Zi-Xin Liu}
\affiliation{School of Physics and Engineering, Zhengzhou University, Zhengzhou 450001, China}

\author{Long Qian}
\affiliation{School of Physics and Engineering, Zhengzhou University, Zhengzhou 450001, China}

\author{Bing Wang}
\affiliation{School of Physics and Engineering, Zhengzhou University, Zhengzhou 450001, China}

\author{Wei Zhang}
\affiliation{School of Physics and Engineering, Zhengzhou University, Zhengzhou 450001, China}

\date{\today}

\begin{abstract}
To study the exotic odd nuclear systems, the self-consistent continuum Skyrme-Hartree-Fock-Bogoliubov theory formulated with Green's function technique is extended to include blocking effects with the equal filling approximation. Detailed formula are presented.
To perform the integrals of the Green's function properly, the contour paths $C_{\rm b}^{-}$ and $C_{\rm b}^{+}$ introduced for the blocking effects should include the blocked quasi-particle state but can not intrude into the continuum area. By comparing with the box-discretized calculations, the great advantages of the Green's function method in describing the extended density distributions, resonant states, and the couplings with the continuum in exotic nuclei are shown. Finally, taking the neutron-rich odd nucleus $^{159}$Sn as an example, the halo structure is investigated by blocking the quasi-particle state $1p_{1/2}$. It is found that it is mainly the weakly bound states near the Fermi surface that contribute a lot for the extended density distributions at large coordinate space.
\end{abstract}

\maketitle
\newcounter{mytempeqncnt}
\section{Introduction}
\label{Sec:Introduction}

With the operation of the worldwide new radioactive ion beam facilities~\cite{NIMPR2002Xia488,NPA2010Zhan834,NPA2010Sturm834,NPA2010Gales834,NPA2010Motobayasgi834,
NPA2010Thoennessen834,KoRIA2010Choi} and the developments in
the detection techniques, the exotic nuclei far from the $\beta$ stability line become a very challenging topic and attract great interests experimentally and theoretically~\cite{AR1993Mueller43,PPNP1995Tanihata35,AP1995Hansen45,PPNP2000Casten45,PhysRadBeams2001Bertulani,
PR2004Jonson389,RMP2004Jensen76,JPG2010Ershov37,SCPMA2011Cao54}. Many new and exotic phenomena such as halos~\cite{PRL1985Tanihata55,PRL1992Minamisono69,ZPA1995Schwab350,PRL1996Meng77,PRL1998MengJ80,PRC2010Zhou82}, changes of nuclear magic numbers~\cite{PRL2000Ozawa84}, and pygmy resonances~\cite{PRL2005Adrich95} have been observed or predicted. In these weakly bound nuclei, the neutron or proton Fermi surface is very close to the continuum threshold, and the valence nucleons can be easily scattered to the continuum due to the pairing correlations. Besides, when the valence nucleon occupy the states with low angular momentum, very extended spatial density distributions as well as large nuclear radius are obtained~\cite{PRL1996Meng77}. As a result, to give a proper theoretical description of the exotic nuclei, one must treat pairing correlations and the couplings with the continuum in a self-consistent way and consider properly the extended asymptotic behavior of nuclear density distributions.

The Hartree-Fock-Bogliubov~(HFB) theory has achieved great successes in describing exotic nuclei with unified description of the mean field and the pairing correlation and properly treatment of the coupling with the continuum. In the spherical case, it has been mainly applied to the Gogny-HFB theory~\cite{PRC1980Decharge21}, Skyrme-HFB theory~\cite{NPA1984Doba422, PRC1996DobaczewskiJ53, PRC2000Mizutori61, PRC2001Grasso64}, the relativistic continuum Hartree-Bogoliubov~(RCHB)~theory~\cite{NPA1998MengJ, PPNP2006MengJ, PhysRep2005Vretenar409, PRL1997Poschl79}, and density dependent relativistic Hartree-Fock-Bogoliubov (RHFB) theory~\cite{PRC2010LongWH81-024308, PRC2010LongWH81-031302, PRC2013LuXL87}.
To describe the halo phenomenon in deformed nuclei, the deformed relativistic Hartree-Bogoliubov (DRHB) theory based on a Woods-Saxon basis~\cite{PRC2010Zhou82,PRC2012LLLI85,CPL2012Li29,PRC2012ChenY85} and the coordinate-space Skyrme-HFB approach with interaction~\cite{PRC2013JCPei87, PRC2013ZhangYN88} have been developed.
Generally, these H(F)B equations can be solved in the coordinate space~\cite{NPA1981Horowitz368} where the Nomerov or Runge-Kutta method ~\cite{CUPC1992Press} can be applied, or in an appropriate basis~\cite{AP1990Gambhir198,PRC1998Stoitsov58,PRC2003Zhou68}. For the exotic nuclei with very extended density distributions, the simple oscillator basis fails due to its localised single-particle wave functions.
Instead, the wave functions in a Woods-Saxon basis have a much more realistic asymptotic behavior at large coordinate. It is shown that the solutions of the relativistic Hartree equations in a Woods-Saxon basis is almost equivalent to the solution in coordinate space~\cite{PRC2003Zhou68}. However, in the spherical systems, compared with the basis expansion method, solving the HFB equations in the coordinate space is more convenient.

In many calculations in the coordinate-space H(F)B approach, the box boundary condition is adopted, and hence the discretized quasiparticle states are obtained~\cite{PRC1996DobaczewskiJ53, PLB2000Bennaceur496, PRL1996Meng77, PPNP2006MengJ}. Although it is appropriate for deeply bound states, the box boundary condition is not suitable for weekly bound and continuum states unless a large enough box is taken. On the other hand, the Green's function method~\cite{YadFiz1987Belyaev45} has a merit to impose the correct asymptotic behaviors on the wave functions especially for the weakly bound and continuum states, and to calculate the densities.

Green's function (GF) method~\cite{PRB1992Tamura_45_3271,PRA2004Foulis_70_022706,Book2006Eleftherios-GF} is an efficient tool in describing continuum, by which the discrete bound states and the continuum can be treated on the same footing; both the energies and widths for the resonant states can be given directly; and the correct asymptotic behaviors for the wave functions can be described. Non-relativistically and relativistically, there are already many applications of the GF method in the nuclear physics to study the contribution of continuum to the ground states and excited states. Non-relativistically, in the spherical case, in 1987, Belyaev \emph{et~al.} constructed the Green's function in the Hartree-Fock-Bogoliubov (HFB) theory in the coordinate representation~\cite{SJNP1987Belyaev_45_783}. Afterwards, Matsuo applied this Green's function to the quasiparticle random-phase approximation (QRPA)~\cite{NPA2001Matsuo696}, which was further used to describe the collective excitations coupled to the continuum~\cite{PTPS2002Matsuo_146_110, PRC2005Matsuo_71_064326,NPA2007Matsuo_788_307,PTP2009Serizawa_121_97,PRC2009Mizuyama_79_024313,PRC2010Matsuo_82_024318,PRC2011Shimoyama_84_044317}, microscopic structures of monopole pair vibrational modes and associated two-neutron transfer amplitudes in neutron-rich Sn isotopes~\cite{PRC2013Shimoyama_88_054308}, and neutron capture reactions in the neutron-rich nuclei~\cite{PRC2015Matsuo_91_034604}. Recently, Zhang \emph{et~al.} developed the fully self-consistent continuum Skyrme-HFB theory with GF method~\cite{PRC2011ZhangY83,PRC2012ZhangY86}. In the deformed case, in 2009, Oba \emph{et~al.} extended the continuum HFB theory to include deformation on the basis of a coupled-channel representation and explored the properties of the continuum and pairing correlation in deformed nuclei near the neutron drip line~\cite{PRC2009Oba80}. Relativistically, in the spherical case, in Refs.~\cite{PRC2009Daoutidis_80_024309,PRC2010DYang_82_054305}, the fully self-consistent relativistic continuum random-phase-approximation (RCRPA) was developed with the Green's function of the Dirac equation and used to study the contribution of the continuum to nuclear collective excitations. In 2014, considering the great successes of the covariant density functional theory~(CDFT)~\cite{ANP1986Sert_16_1,PPNP2006MengJ,JPG2015MengJ_42_093101,PRC2013Zhang_88_054324,CPC2017Zhang,PRC2017Zhang_96_054308,PRC2018Zhang_97_054302,PRC2016TTSun,JPG2017Lu_44_125104,PRC2017TTSun_96_044312,CPC2018Sun,
PRD2018Sun_98_034031,PRDSun_99_023004}, the authors developed the continuum CDFT based on the GF method, with which the accurate energies and widths of the single-neutron resonant states were calculated for the first time~\cite{PRC2014TTSun_90_054321}. This method has been further extended to describe single-particle resonances for protons~\cite{JPG2016TTSun_43_045107} and $\Lambda$ hyperons~\cite{PRC2017Ren_95_054318}. In 2016, further containing pairing correlation, the Green's function relativistic continuum Hartree-Bogoliubov (GF-RCHB) theory was developed, by which the continuum was treated exactly and the giant halo phenomena in neutron-rich Zr isotopes were studied~\cite{Sci2016Sun_46_12006}.

However, the above Skyrme HFB theory with the Green's function method is only formulated for even-even nuclei~\cite{Shlomo1975507,NPA2001Matsuo696,PRC2009Oba80, PRC2011ZhangY83, PRC2012ZhangY86}.
To describe the exotic nuclear structure in neutron rich odd-$A$ nuclei, the blocking effect has to be taken into account.
In the work, we extend the continuum Skyrme-HFB theory with Green's function method to discuss odd-$A$ nuclei by incorporating the blocking effect. In this way, pairing correlations, continuum, blocking effects can be described consistently in the coordinate space.

The paper is organized as follows: In Sec.~\ref{Sec:Theory}, we introduce the formulation of the continuum Skyrme-HFB theory for odd-$A$ nuclei using the Green's function technique. Numerical details and checks will be presented in Sec.~\ref{Sec:Numerical}. After giving the results and discussions in Sec.~\ref{Sec:Results}, finally conclusions are drawn in Sec.~\ref{Sec:Summary}.

\section{THEORETICAL FRAMEWORK}
\label{Sec:Theory}

\subsection{Coordinate-space Hartree-Fock-Bogoliubov theory}

In the Hartree-Fock-Bogoliubov (HFB) theory, the pair correlated nuclear system is described in terms of independent quasiparticles~\cite{ManybodyProb2000}. In the coordinate space, the HFB equation for the quasiparticle state $\phi_{i}(\bm{r}\sigma)$ is written as~\cite{NPA1984Doba422}
\begin{equation}
\left(
   \begin{array}{cc}
     h-\lambda    & \tilde{h}\\
    \tilde{h}^{*} & -h^{*}+\lambda \\
   \end{array}
\right)
\phi_{i}(\bm{r}\sigma)
=E_{i}\phi_{i}(\bm{r}\sigma),
\label{EQ:HFBeq}
\end{equation}%
with the quasi-particle energy $E_{i}$ and the Fermi energy $\lambda$ determined by constraining the expectation value of the nucleon number. The solutions of HFB equation have two symmetric branches. One is positive ($E_i>0$) with wave function $\phi_{i}(\bm{r}\sigma)$, and the other one is negative ($-E_i<0$) with conjugate wave function $\bar{\phi}_{\tilde{i}}(\bm{r}\sigma)$. The quasi-particle wave function $\phi_{i}(\bm{r}\sigma)$ and its conjugate wave function $\bar{\phi}_{\tilde{i}}(\bm{r}\sigma)$ have two components,
\begin{equation}
 \phi_{i}(\bm{r}\sigma)\equiv
   \left(
     \begin{array}{c}
       \varphi_{1,i}(\bm{r}\sigma) \\
       \varphi_{2,i}(\bm{r}\sigma) \\
     \end{array}
   \right),~~~~
 \bar{\phi}_{\tilde{i}}(\bm{r}\sigma)\equiv
   \left(
     \begin{array}{c}
       -\varphi_{2,i}^{*}(\bm{r}\tilde{\sigma}) \\
       ~~~\varphi_{1,i}^{*}(\bm{r}\tilde{\sigma}) \\
     \end{array}
  \right),
  \label{EQ:qpwf}
\end{equation}
where $\varphi(\bm{r}\tilde{\sigma})\equiv -2\sigma\varphi(\bm{r}, -\sigma)$. Note that the notations in this paper follow Ref.~\cite{NPA2001Matsuo696}.
The Hartree-Fock hamiltonian $h(\bm{r}\sigma, \bm{r'}\sigma')$ and the pair hamiltonian $\tilde{h}(\bm{r}\sigma, \bm{r'}\sigma')$ can be respectively obtained by the variation of the total energy functional with respect to the particle density $\rho(\bm{r}\sigma, \bm{r}'\sigma')$ and pair density $\tilde{\rho}(\bm{r}\sigma, \bm{r}'\sigma')$,
\begin{subequations}
\begin{eqnarray}
\rho(\bm{r}\sigma, \bm{r}'\sigma')&\equiv& \langle\Phi_{0}|c^{\dag}_{\bm{r}'\sigma'}c_{\bm{r}\sigma}|\Phi_{0}\rangle,\\
\tilde{\rho}(\bm{r}\sigma, \bm{r}'\sigma')&\equiv& \langle\Phi_{0}|c_{\bm{r}'\tilde{\sigma}'}c_{\bm{r}\sigma}|\Phi_{0}\rangle,
\end{eqnarray}
\end{subequations}
where $|\Phi_{0}\rangle$ is the ground state of the system, $c_{\bm{r}\sigma}$ and $c_{\bm{r}\sigma}^{\dag}$ are the particle annihilate and creation operators, respectively. The two density matrices can be combined in a generalized density matrix $R$ as
\begin{eqnarray}
&& R(\bm{r}\sigma,\bm{r'}\sigma')\nonumber\\
&\equiv& \left(
   \begin{array}{cc}
     \rho(\bm{r}\sigma, \bm{r}'\sigma')
    &\tilde{\rho}(\bm{r}\sigma, \bm{r}'\sigma') \\
     \tilde{\rho}^{*}(\bm{r}\tilde{\sigma}, \bm{r}'\tilde{\sigma}')
    &\delta_{{\bm r}{\bm r'}}\delta_{\sigma\sigma'}-\rho^{*}(\bm{r}\tilde{\sigma}, \bm{r}'\tilde{\sigma}')
   \end{array}
  \right),
  \label{EQ:Rmarix}
\end{eqnarray}
where the particle density $\rho(\bm{r}\sigma, \bm{r}'\sigma')$ and pair density $\tilde{\rho}(\bm{r}\sigma, \bm{r}'\sigma')$ are the $``11"$ and $``12"$ components of $R(\bm{r}\sigma,\bm{r'}\sigma')$, respectively.

For an even-even nucleus, the ground state $|\Phi_{0}\rangle$ is represented as a vacuum with
respect to quasiparticles~\cite{ManybodyProb2000}, i.e.,
\begin{equation}
 \beta_{i}|\Phi_{0}\rangle=0,~~{\rm for~all}~~i=1,\cdots, M,
\end{equation}
where $\beta_{i}$ and $\beta_{i}^{\dag}$ are the quasiparticle annihilation and creation operators which are obtained by the Bogoliubov transformation from the particle operators $c_{\bm{r}\sigma}$ and $c_{\bm{r}\sigma}^{\dag}$, and $M$ is the dimension of the quasiparticle space.

Starting from the bare vacuum $|0\rangle$, the ground state $|\Phi_{0}\rangle$ for an even-even nucleus can be constructed as,
\begin{equation}
|\Phi_0\rangle =\prod_{i}\beta_i |0\rangle,
\end{equation}
where $i$ runs over all values of $i=1,2,\cdots, M$.

With the quasiparticle vacuum $|\Phi_{0}\rangle$, the generalized density matrix can be expressed in a simple form,
\begin{equation}
R(\bm{r}\sigma,\bm{r}'\sigma')=\sum\limits_{i:{\rm all}}\bar{\phi}_{\tilde{i}}(\bm{r}\sigma)\bar{\phi}^{\dag}_{\tilde{i}}(\bm{r}'\sigma').
\label{EQ:Reven}
\end{equation}

\subsection{Blocking effect for odd-$A$ nuclei}

For an odd-$A$ nucleus£¬the ground state is a one-quasiparticle state~$|\Phi_{1}\rangle$~\cite{ManybodyProb2000}, which can be constructed based on a HFB vacuum~$|\Phi_{0}\rangle$ as
\begin{equation}
   |\Phi_{1}\rangle=\beta_{i_{\rm b}}^{\dag}|\Phi_{0}\rangle,
   \label{EQ:WFodd}
\end{equation}
where $i_{\rm b}$ denotes the blocked quasi-particle state occupied by the odd nucleon. For the ground state of the odd system, $\beta_{i_{\rm b}}=\beta_1$ corresponds to the quasi-particle state with the lowest quasi-particle energy. The state $|\Phi_{1}\rangle$ is a vacuum to the operators $(\tilde{\beta}_{i_{\rm b}}, \tilde{\beta}_2, \cdots, \tilde{\beta}_M)$ with
\begin{equation}
\tilde{\beta}_{i_{\rm b}}=\beta_{1}^{\dag},\tilde{\beta}_2=\beta_{2},\cdots,\tilde{\beta}_M=\beta_{M},
\end{equation}
where the exchange of the operators $\beta_{i_{\rm b}}^{\dag}\leftrightarrow\beta_{i_{\rm b}}$ (or $\beta_{1}^{\dag}\leftrightarrow\beta_{1}$) corresponds to the exchange of the wave function
\begin{equation}
\phi_{i_{\rm b}}(\bm{r}\sigma)\leftrightarrow\bar{\phi}_{\tilde{i}_{\rm b}}(\bm{r}\sigma).
\end{equation}
Accordingly, the particle density $\rho(\bm{r}\sigma,\bm{r}'\sigma')$ and pair density $\tilde{\rho}(\bm{r}\sigma,\bm{r}'\sigma')$ for the one-quasiparticle state $|\Phi_{1}\rangle$ are,
\begin{subequations}
 \begin{eqnarray}
  \rho(\bm{r}\sigma,\bm{r}'\sigma')&\equiv\langle\Phi_{1}|c^{\dag}_{\bm{r}'\sigma'}c_{\bm{r}\sigma}|\Phi_{1}\rangle,\\
  \tilde{\rho}(\bm{r}\sigma,\bm{r}'\sigma')&\equiv\langle\Phi_{1}|c_{\bm{r}'\tilde{\sigma}'}c_{\bm{r}\sigma}|\Phi_{1}\rangle,
 \end{eqnarray}
\end{subequations}
and the generalized density matrix~$R(\bm{r}\sigma,\bm{r}'\sigma')$ becomes
\begin{eqnarray}
&&R(\bm{r}\sigma,\bm{r}'\sigma')=\sum\limits_{i:{\rm all}}\bar{\phi}_{\tilde{i}}(\bm{r}\sigma)\bar{\phi}^{\dag}_{\tilde{i}}(\bm{r}'\sigma')~~~~~\nonumber\\
&&
~~~~~~~~~~~~~~~
-\bar{\phi}_{\tilde{i}_{\rm b}}(\bm{r}\sigma)\bar{\phi}^{\dag}_{\tilde{i}_{\rm b}}(\bm{r}'\sigma')
+\phi_{i_{\rm b}}(\bm{r}\sigma)\phi_{i_{\rm b}}^{\dag}(\bm{r}'\sigma'),
 \label{EQ:Rodd}
\end{eqnarray}
where two more terms are introduced compared with those for even-even nuclei in Eq.~(\ref{EQ:Reven}) after including the blocking effect in odd nuclear systems.

\subsection{Density and quasi-particle spectrum}

In the conventional Skyrme-HFB theory, one solves the HFB equation (\ref{EQ:HFBeq}) with the box boundary condition to obtain the discretized eigensolutions for the single-quasiparticle energy and the corresponding wave functions. Then the generalized density matrix $R({\bm r}\sigma,{\bm r'}\sigma')$ can be constructed by a sum over discretized quasiparticle states. We call this method as box-discretized Skyrme-HFB approach. However, the box boundary condition is not appropriate for the description of weakly bound states and continuum in exotic nuclei unless a large enough box size is taken.

Instead, Green's function method is used to impose the correct asymptotic behaviors on the wave functions especially for the continuum states, and to calculate the densities. The Green's function $G({\bm r}\sigma,{\bm r'}\sigma';E)$ defined for the coordinate-space HFB equation obeys,
\begin{eqnarray}
\left[E-
\left(
  \begin{array}{cc}
    h-\lambda & \tilde{h} \\
    \tilde{h}^* & -h^*+\lambda \\
  \end{array}
\right)
\right]G({\bm r}\sigma,{\bm r'}\sigma';E)&&\nonumber\\
=\delta({\bm r}-{\bm r'})\delta_{\sigma\sigma'}.&&
\label{EQ:HFBGF-define}
\end{eqnarray}
With a complete set of eigenstates $\{\phi_{i}({\bm r}\sigma),\bar{\phi}_{\tilde{i}}({\bm r}\sigma)\}$ and eigenvalues $\{E_{i},-E_i\}$ of the HFB equation, the HFB Green's function in Eq.~(\ref{EQ:HFBGF-define}) can be represented as
\begin{eqnarray}
 & &G({\bm r}\sigma, {\bm r}'\sigma';E)\nonumber\\
 &=&\sum\!\!\!\!\!\!\!\!\!\int\left(\frac{\phi_{i}({\bm r}\sigma)\phi_{i}^{\dag}({\bm r}'\sigma')}{E-E_{i}}+\frac{\bar{\phi}_{\tilde{i}}({\bm r}\sigma)\bar{\phi}_{\tilde{i}}^{\dag}({\bm r}'\sigma')}{E+E_{i}}\right),
 \label{EQ:GF}
\end{eqnarray}
which has two branches. One is for $\phi_{i}({\bm r}\sigma)$ and $E_{i}$, and the other is for $\bar{\phi}_{\tilde{i}}({\bm r}\sigma)$ and $-E_{i}$. The $\sum\!\!\!\!\!\!\!\int$ is summation for the quasi-particle discrete states with the quasi-particle energy $|E_{i}|<|\lambda|$ and integral for the continuum with $|E_{i}|>|\lambda|$ explicitly.

Corresponding to the upper and lower components of the quasi-particle wave function,
the Green's function for the HFB equation can be written as a $2\times 2$ matrix,
\begin{equation}
G({\bm r}\sigma,{\bm r'}\sigma';E)=
\left(
  \begin{array}{cc}
    G^{(11)}(E) & G^{(12)}(E) \\
    G^{(21)}(E) & G^{(22)}(E) \\
  \end{array}
\right).
\end{equation}

\begin{figure}[!t]
  \includegraphics[width=0.45\textwidth]{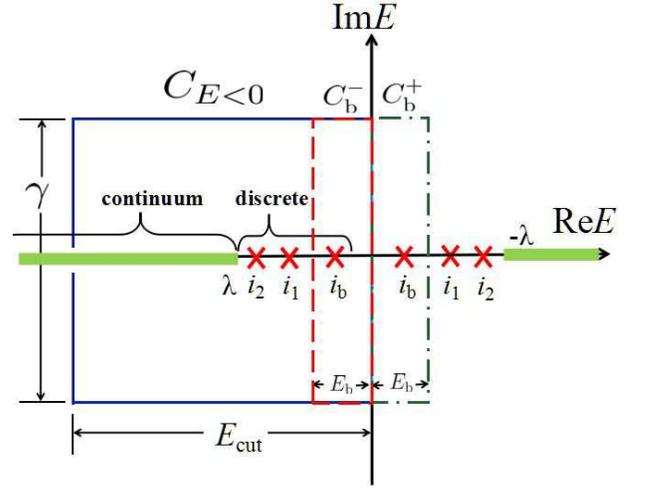}\\
  \caption{Contour paths $C_{E<0}, C_{\rm b}^{-}, C_{\rm b}^{+}$ are to perform the integrations of the Green's function on the complex quasiparticle energy plane. The paths are chosen to be rectangles with the same width $\gamma$ and different lengths, i.e., $E_{\rm cut}$, $E_{\rm b}$, and $E_{\rm b}$ for $C_{E<0}, C_{\rm b}^{-}$, and $C_{\rm b}^{+}$ respectively. The crosses denote the discrete quasiparticle states and the continuum states are denoted by the thick solid line. Quantum number $i_{\rm b}$ denotes the single quasi-particle state to be blocked.}
  \label{Fig1}
\end{figure}

Starting from Eq.~(\ref{EQ:GF}) and according to the Cauchy's theorem, the generalized density matrix in Eq.~(\ref{EQ:Rodd}) can be calculated with the integrals of the Green's function in the complex quasiparticle energy plane as
\begin{eqnarray}
\label{Eq:density-GF}
&&R({\bm r}\sigma,{\bm r}'\sigma')=\frac{1}{2\pi i}\left[\oint_{C_{E<0}} dE G({\bm r}\sigma,{\bm r}'\sigma';E)\right.\\
&&~~~~~~~~
\left.-\oint_{C_{\rm b}^{-}} dE G({\bm r}\sigma,{\bm r}'\sigma';E)+\oint_{C_{\rm b}^{+}} dE G({\bm r}\sigma,{\bm r}'\sigma';E)\right],\nonumber
\end{eqnarray}
where the contour path $C_{E<0}$ encloses all the negative quasiparticle energies $-E_{i}$, $C_{\rm b}^{-}$ encloses only the pole $-E_{i_{\rm b}}$ and $C_{\rm b}^{+}$ encloses only the pole $E_{i_{\rm b}}$, which can be seen in Fig.~\ref{Fig1}. Note that the three terms in Eq.~(\ref{Eq:density-GF})
corresponds one-to-one to those in Eq.~(\ref{EQ:Rodd}).

In the spherical case, the quasiparticle wave function $\phi_{i}({\bm r}\sigma)$ and the conjugate wave function $\bar{\phi}_{\tilde{i}}({\bm r}\sigma)$ can be expanded as
\begin{subequations}
\begin{eqnarray}
&&\phi_i({\bm r}\sigma)=\frac{1}{r}\phi_{nlj}(r)Y_{jm}^l(\hat{\bm r}\sigma),~~
\phi_{nlj}(r)=
\left(
  \begin{array}{c}
    \varphi_{1,nlj}(r) \\
    \varphi_{2,nlj}(r) \\
  \end{array}
\right),
\label{EQ:rWF}\\
&&\bar{\phi}_{\tilde{i}}({\bm r}\sigma)=\frac{1}{r}\bar{\phi}_{nlj}(r)Y_{jm}^{l*}(\hat{\bm r}\tilde{\sigma}),~~
\bar{\phi}_{nlj}(r)=
\left(
  \begin{array}{c}
    -\varphi_{2,nlj}^{*}(r) \\
    ~~~\varphi_{1,nlj}^*(r) \\
  \end{array}
\right),~~~~~~~~
\end{eqnarray}
\end{subequations}%
where $Y_{jm}^l(\hat{\bm r}\sigma)$ is the spin spherical harmonic, and $Y_{jm}^l(\hat{\bm r}\tilde{\sigma})=-2\sigma Y_{jm}^l(\hat{\bm r}-\sigma)$. Similarly, the generalized density matrix $R({\bm r}\sigma,{\bm r}'\sigma')$ and the Green's function $G({\bm r}\sigma,{\bm r}'\sigma';E)$
can also be expanded as
\begin{subequations}%
\begin{eqnarray}
 R({\bm r}\sigma, {\bm r'}\sigma')&=&\sum_{ljm}Y_{jm}^l(\hat{\bm r}\sigma)R_{lj}(r,r')Y_{jm}^{l*}(\hat{\bm r}'\sigma'),\\
G({\bm r}\sigma, {\bm r'}\sigma';E)&=&\sum_{ljm}Y_{jm}^l(\hat{\bm r}\sigma)\frac{\mathcal{G}_{lj}(r,r';E)}{rr'}Y_{jm}^{l*}(\hat{\bm r}'\sigma'),~~~~~~~~%
\end{eqnarray}%
\end{subequations}%
where $R_{lj}(r,r')$ and $\mathcal{G}_{lj}(r,r';E)$ are the radial parts of the generalized density matrix and Green's function, respectively. Note that the equal filling approximation is applied for the odd nucleon, i.e., we take an average of the blocked quasiparticle state $i_{\rm b}=(n_{\rm b}l_{\rm b}j_{\rm b}m_{j_{\rm b}})$ over the magnetic quantum numbers $m_{j_{\rm b}}= -j_{\rm b}, -j_{\rm b}+1, \cdots, j_{\rm b}-1, j_{\rm b}$.

As a result, the radial local generalized density matrix $R(r)=R(r,r)$
can be expressed by the radial box-discretized quasiparticle wave functions $\phi_{nlj}(r)$ or the radial HFB Green's function $\mathcal{G}_{lj}(r,r';E)$ as
\begin{eqnarray}
&&R(r)=\sum_{lj}R_{lj}(r,r)\nonumber\\
&=&\frac{1}{4\pi r^{2}}\left[\sum_{lj:{\rm all}}(2j+1)\sum_{n:{\rm all}}\bar{\phi}_{nlj}^{2}(r)-\bar{\phi}_{n_{\rm b}l_{\rm b}j_{\rm b}}^{2}(r)+\phi_{n_{\rm b}l_{\rm b}j_{\rm b}}^{2}(r)\right]\nonumber\\
&=&\frac{1}{4\pi r^{2}}\frac{1}{2\pi i}\left[\sum_{lj:{\rm all}}(2j+1)\oint_{C_{E<0}}dE\mathcal{G}_{lj}(r,r;E)\right.\nonumber\\
&&~~~~\left.-\oint_{C^{-}_{\rm b}}dE\mathcal{G}_{l_{\rm b}j_{\rm b}}(r,r;E)+\oint_{C^{+}_{\rm b}}dE\mathcal{G}_{l_{\rm b}j_{\rm b}}(r,r;E)\right].%
\label{EQ:Rr}%
\end{eqnarray}%
From the radial generalized matrix $R(r)$, one can easily obtain the radial local particle density $\rho(r)$ and pair density $\tilde{\rho}(r)$, which are the ``11" and ``12" components of $R(r)$, respectively. In the same way, one can express other radial local densities needed in the functional of the Skyrme interaction~\cite{NPA1975Engel249, RevModPhys2003Bender75}, such as the kinetic-energy density $\tau(r)$, the spin-orbit density $J(r)$, and etc., in terms of the radial Green's function.

Accordingly, the particle density and pair density for the blocked partial wave $lj=l_{\rm b}j_{\rm b}$ can be written as,
\begin{subequations}
\begin{eqnarray}
&&\rho_{lj}(r)=\rho_{0,lj}(r)-\rho_{1,lj}(r)+\rho_{2,lj}(r)\nonumber\\
&&=\frac{1}{4\pi r^{2}}\left[(2j+1)\sum_{n}\varphi^{2}_{2,nlj}(r)-\varphi^{2}_{2,n_{\rm b}l_{\rm b}j_{\rm b}}(r)+\varphi^{2}_{1,n_{\rm b}l_{\rm b}j_{\rm b}}(r)\right]\nonumber\\
&&=\frac{1}{4\pi r^{2}}\frac{1}{2\pi i}\left[(2j+1)\oint_{C_{E}<0}dE \mathcal{G}^{(11)}_{lj}(r,r;E) \right.\nonumber\\
&&~~~~~~~\left.-\oint_{C^{-}_{\rm b}}dE\mathcal{G}^{(11)}_{lj}(r,r;E)+\oint_{C^{+}_{\rm b}}dE\mathcal{G}^{(11)}_{lj}(r,r;E)\right];
\label{EQ:rholj}
\\
&&\tilde{\rho}_{lj}(r)=\tilde{\rho}_{0,lj}(r)-\tilde{\rho}_{1,lj}(r)+\tilde{\rho}_{2,lj}(r)\nonumber\\
&&=\frac{1}{4\pi r^{2}}\left[(2j+1)\sum_{n}\varphi_{1,nlj}(r)\varphi_{2,nlj}(r)\right.\nonumber\\
&&~~~~\left.-\varphi_{1,n_{\rm b}l_{\rm b}j_{\rm b}}(r)\varphi_{2,n_{\rm b}l_{\rm b}j_{\rm b}}(r)
-\varphi_{2,n_{\rm b}l_{\rm b}j_{\rm b}}(r)\varphi_{1,n_{\rm b}l_{\rm b}j_{\rm b}}(r)
\right]\nonumber\\
&&=\frac{1}{4\pi r^{2}}\frac{1}{2\pi i}\left[(2j+1)\oint_{C_{E}<0}dE \mathcal{G}^{(12)}_{lj}(r,r;E) \right.\nonumber\\
&&~~~~~\left.-\oint_{C^{-}_{\rm b}}dE\mathcal{G}^{(12)}_{lj}(r,r;E)+\oint_{C^{+}_{\rm b}}dE\mathcal{G}^{(12)}_{lj}(r,r;E)\right].
\label{EQ:trholj}
\end{eqnarray}
\end{subequations}
And for the partial waves with $lj\neq l_{\rm b}j_{\rm b}$, the terms introduced by blocking effect, i.e., $\rho_{1,lj}(r)$ and $\rho_{2,lj}(r)$ in $\rho_{lj}(r)$, and $\tilde{\rho}_{1,lj}(r)$ and $\tilde{\rho}_{2,lj}(r)$ in $\tilde{\rho}_{lj}(r)$, are zero.

Within the framework of the continuum Skyrme-HFB theory, the quasi-particle energy spectrum can be given by the occupation number density $n_{lj}(E)$ or the pair number density $\tilde{n}_{lj}(E)$. The integrals of them with energy $E$ represent the occupied nucleon number $N_{lj}$ and paired nucleon number $\tilde{N}_{lj}$ in partial wave $lj$, i.e.,
\begin{subequations}
\begin{eqnarray}
          N_{lj}&=&\int dEn_{lj}(E),\\
  \tilde{N}_{lj}&=&\int dE\tilde{n}_{lj}(E).
\end{eqnarray}
\end{subequations}

For odd-$A$ nuclei, the occupation number density $n_{lj}(E)$ and the pair number density $\tilde{n}_{lj}(E)$ for the partial wave $lj=l_{\rm b}j_{\rm b}$ can be written as
\begin{subequations}
\begin{eqnarray}
n_{lj}(E) &=& n_{0,lj}(E)-n_{1,lj}(E)+n_{2,lj}(E)\nonumber\\
&=&\frac{2j+1}{\pi}\int dr\textbf{Im}\mathcal{G}^{(11)}_{0,lj}(r,r;-E-i\epsilon){\Big|}_{-E=-E_{\rm cut}}^{0}\nonumber\\
&&-\frac{1}{\pi}\int dr \textbf{Im}\mathcal{G}^{(11)}_{l_{\rm b}j_{\rm b}}(r,r;-E-i\epsilon){\Big|}_{-E=-E_{\rm b}}^{0}
\nonumber\\
&&+\frac{1}{\pi}\int dr \textbf{Im}\mathcal{G}^{(11)}_{l_{\rm b}j_{\rm b}}(r,r;E-i\epsilon){\Big|}_{E=0}^{E_{\rm b}};\label{EQ:nE}\\%
\tilde{n}_{lj}(E) &=&\tilde{n}_{0,lj}(E)-\tilde{n}_{1,lj}(E)+\tilde{n}_{2,lj}(E)\nonumber\\
&=&\frac{2j+1}{\pi}\int dr\textbf{Im}\mathcal{G}^{(12)}_{0,lj}(r,r;-E-i\epsilon){\Big|}_{-E=-E_{\rm cut}}^{0}\nonumber\\
&&-\frac{1}{\pi}\int dr \textbf{Im}\mathcal{G}^{(12)}_{l_{\rm b}j_{\rm b}}(r,r;-E-i\epsilon){\Big|}_{-E=-E_{\rm b}}^{0}
\nonumber\\
&&+\frac{1}{\pi}\int dr \textbf{Im}\mathcal{G}^{(12)}_{l_{\rm b}j_{\rm b}}(r,r;E-i\epsilon){\Big|}_{E=0}^{E_{\rm b}},%
\label{EQ:ntE}%
\end{eqnarray}%
\end{subequations}%
where the terms $n_{1,lj}(E)$ and $n_{2,lj}(E)$ in $n_{lj}(E)$ and $\tilde{n}_{1,lj}(E)$ and $\tilde{n}_{2,lj}(E)$ in $\tilde{n}_{lj}(E)$ are introduced due to the blocking effect and they are zero for the partial waves with $lj\neq l_{\rm b}j_{\rm b}$. The energy ranges of the Green's functions in the terms $n_{0,lj}(E)$, $n_{1,lj}(E)$, and $n_{2,lj}(E)$ are $-E_{\rm cut}<-E<0$, $-E_{\rm b}<-E<0$ and $0<E<E_{\rm b}$, which are in accordance with the real energy ranges of the contour paths $C_{E<0}$, $C^{-}_{\rm b}$ and $C^{+}_{\rm b}$ in Fig.~\ref{Fig1}.

\subsection{Construction of HFB Green's function}
For given quasi-particle energy $E$ and quantum number $lj$, the radial HFB Green's function $\mathcal{G}_{lj}(r,r';E)$ can be constructed as
\begin{eqnarray}
\label{EQ:rGF}
&&\mathcal{G}_{lj}(r,r';E)=\sum_{s,s'=1,2}c^{ss'}_{lj}\left[\theta(r-r')\phi_{lj}^{(+s)}(r,E)\phi_{lj}^{(rs')T}(r',E)\right.\nonumber\\
&&~~~~~~~~~~~~~~~~~~~~~~
\left.+\theta(r'-r)\phi_{lj}^{(rs')}(r,E)\phi_{lj}^{(+s)T}(r',E)\right],
\end{eqnarray}
where $\theta(r-r')$ is the step function, $\phi_{lj}^{(rs)}(r,E)$ and $\phi_{lj}^{(+s)}(r,E)$ $(s=1,2)$ are independent solutions of the radial HFB equation,
\begin{equation}
\phi_{lj}^{(rs)}(r,E)=\left(
                        \begin{array}{c}
                          \varphi_{1,lj}^{(rs)}(r,E) \\
                          \varphi_{2,lj}^{(rs)}(r,E) \\
                        \end{array}
                      \right),
\phi_{lj}^{(+s)}(r,E)=\left(
                        \begin{array}{c}
                          \varphi_{1,lj}^{(+s)}(r,E) \\
                          \varphi_{2,lj}^{(+s)}(r,E) \\
                        \end{array}
                      \right),
\end{equation}
obtained by Runge-Kutta integral starting from the boundary conditions at the origin, $r=0$, and at the edge of the box, $r=R$, respectively. The coefficients $c^{ss'}_{lj}(E)$ are expressed in terms of the Wronskians as
\begin{equation}
\left(
   \begin{array}{cc}
     c^{11}_{lj}    & c^{12}_{lj}\\
     c^{21}_{lj}    & c^{22}_{lj} \\
   \end{array}
\right)=
\left(
   \begin{array}{cc}
     w_{lj}(r1,+1)    & w_{lj}(r1,+2)\\
     w_{lj}(r2,+1)    & w_{lj}(r2,+2) \\
   \end{array}
\right)^{-1},
\end{equation}
with
\begin{eqnarray}
&&w_{lj}(rs,+s')\nonumber\\
&=& \frac{\hbar^{2}}{2m}\left[\varphi^{(rs)}_{1,lj}(r)\frac{d}{dr}\varphi^{(+s')}_{1,lj}(r)-
\varphi^{(+s')}_{1,lj}(r)\frac{d}{dr}\varphi^{(rs)}_{1,lj}(r)\right.\nonumber\\
&~&\left.-\varphi^{(rs)}_{2,lj}(r)\frac{d}{dr}\varphi^{(+s')}_{2,lj}(r)+
 \varphi^{(+s')}_{2,lj}(r)\frac{d}{dr}\varphi^{(rs)}_{2,lj}(r)\right].
\end{eqnarray}

To impose the correct asymptotic behavior on the wave function for the continuum states, we adopt the boundary condition as follows,
\begin{equation}
\left\{
\begin{array}{lcl}
\phi^{(rs)}_{lj}(r,E):&&{\rm regular~at~the~origin}~r=0 \\
\phi^{(+s)}_{lj}(r,E):&&{\rm outgoing~wave~at}~{\displaystyle r \rightarrow\infty}
\end{array}
\right.,
\end{equation}

Explicitly, the solutions $\phi^{(+s)}_{lj}(r,E)$ at $r>R$ satisfy
\begin{equation}
\phi^{(+1)}_{lj}(r,E)\rightarrow\left(
   \begin{array}{c}
   e^{ik_{+}(E)r}\\
   0
   \end{array}
   \right),
\phi^{(+2)}_{lj}(r,E)\rightarrow\left(
   \begin{array}{c}
   0\\
   e^{ik_{-}(E)r}
   \end{array}
   \right).
\end{equation}
Here $k_{\pm}(E)=\sqrt{2m(\lambda\pm E)}/\hbar$ with $m$ the nucleon mass and their branch cuts are
chosen so that ${\rm Im}k_{\pm}>0$ is satisfied.

\begin{figure*}[!t]
 \includegraphics[width=0.8\textwidth]{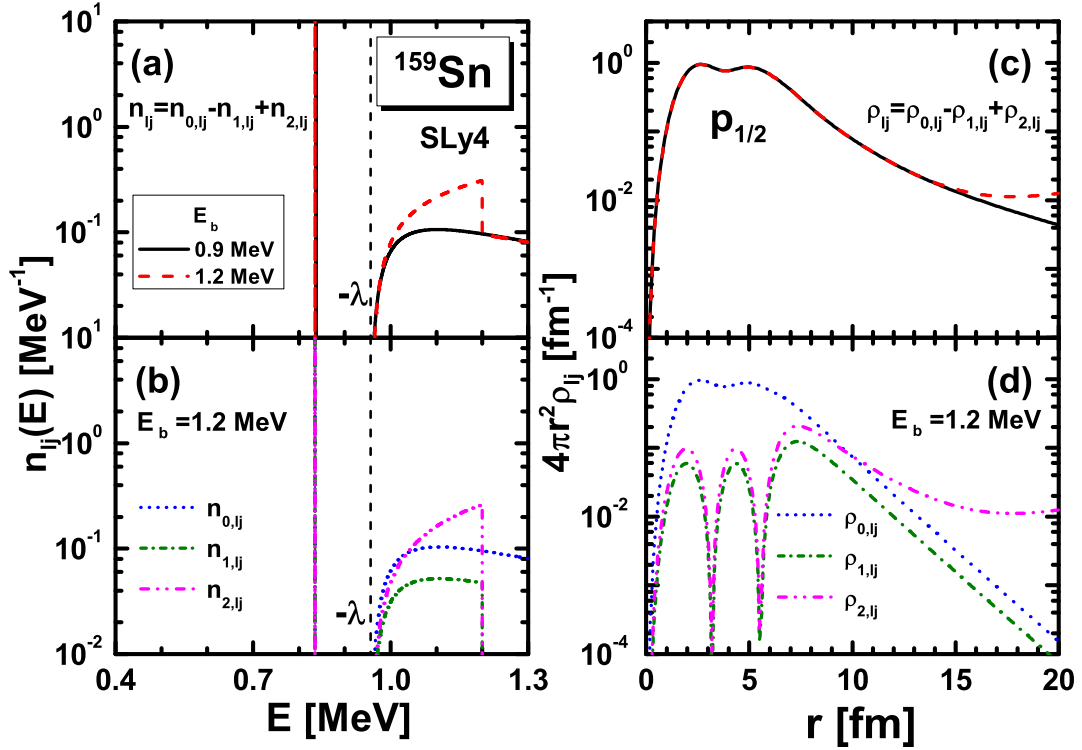}
 \caption{(a) Neutron occupation number density $n_{lj}(E)$ around the continuum threshold energy $-\lambda$ and (b) neutron density distributions $4\pi r^{2}\rho_{lj}(r)$ for the partial wave $p_{1/2}$ of $^{159}$Sn obtained in the continuum Skyrme HFB calculation by blocking quasiparticle state $3p_{1/2}$ with different blocking energy widths $E_{\rm b}=0.9$~MeV (solid lines) and $1.2$~MeV (dashed lines). For the case of the blocking energy width $E_{\rm b}>-\lambda$, i.e., $E_{\rm b}=1.2$ MeV, (b) the contributions for $n_{lj}(E)$ from the terms $n_{0,lj}(E)$, $n_{1,lj}(E)$, and $n_{2,lj}(E)$ in Eq.~(\ref{EQ:nE}), and (d) the contributions for $\rho_{lj}(r)$ from the terms $\rho_{0,lj}(r)$, $\rho_{1,lj}(r)$, and $\rho_{2,lj}(r)$ in Eq.~(\ref{EQ:rholj}) are presented.}
 \label{Fig2}
\end{figure*}
\section{NUMERICAL DETAILS AND CHECKS}
\label{Sec:Numerical}
In this part, numerical details and checks in the continuum Skyrme-HFB calculations are presented for odd nuclear systems.
Besides, the advantages of the Green's function method are shown compared with those by discretized method with box boundary conditions.

\subsection{Numerical details}
In the $ph$ channel, the Skyrme parameter SLy4~\cite{NPA1998ChabanatM635Skyrme} is taken. In the $pp$ channel, a density dependent $\delta$ interaction (DDDI) is adopted for the pairing interaction,
\begin{equation}
v_{\rm pair}({\bm r},{\bm r'})=\frac{1}{2}(1-P_{\sigma})V_0\left[1-\eta\left(\frac{\rho({\bm r})}{\rho_0}\right)^{\alpha}\right]\delta({\bm r}-{\bm r'}),
 \label{EQ:DDDI}
\end{equation}
with which the pair Hamiltonian $\tilde{h}({\bm r}\sigma, {\bm r}'\sigma')$ is reduced to a local pair potential~\cite{NPA1984Doba422}
\begin{equation}
 \Delta({\bm r})=\frac{1}{2}V_0\left[1-\eta\left(\frac{\rho({\bm r})}{\rho_0}\right)^\alpha\right]\tilde{\rho}({\bm r}),
\end{equation}
where $\rho({\bm r})$ and $\tilde{\rho}({\bm r})$ are the particle density and pair density, respectively. The parameters in DDDI are taken as $V_{0}=-458.4$~MeV$\cdot$fm$^{3}$, $\eta=0.71$, $\alpha=0.59$, and $\rho_{0}=0.08$~fm$^{-3}$, which are constrained by reproducing the experimental neutron pairing gaps for the Sn isotopes~\cite{PRC2006MatsuoM73, NPA2007Matsuo788, PRC2010Matsuo82} and the scattering length $a=-18.5$~fm in the $^{1}S$ channel of the bare nuclear force in the low density limit~\cite{PRC2006MatsuoM73}. The cut-off of the quasiparticle states are taken with maximal angular momentum $j_{\rm max}=25/2$ and the maximal quasiparticle energy $E_{\rm cut}=60$~MeV.

To perform the integrals of the Green's function, the contour paths $C_{E<0}, C^{-}_{\rm b}, C^{+}_{\rm b}$ are chosen to be three rectangles on the complex quasiparticle energy plane as shown in Fig.~\ref{Fig1}, with the same height $\gamma=0.1$~MeV and different widths, i.e., $E_{\rm cut}$, $E_{\rm b}$, $E_{\rm b}$ respectively~\cite{PRC2011ZhangY83}. To enclose all the negative quasiparticle energies, the length of the contour path $C_{E<0}$ is taken as the maximal quasiparticle energy $E_{\rm cut}=60$~MeV. The contour paths $C^{+}_{\rm b}$ and $C^{-}_{\rm b}$ are symmetric with respect to the origin and have the same length $E_{\rm b}$, which enclose the blocked quasiparticle states at $E_{i_{\rm b}}$ and $-E_{i_{\rm b}}$, respectively. For the contour integration, we adopt an energy step $\Delta E=0.01$~MeV on the contour path. The HFB equation is solved with the box size $R=20$~fm and mesh size $\Delta r=0.1$~fm in the coordinate space.

\subsection{Numerical checks}
In the following, taking the odd-even neutron-rich nucleus $^{159}$Sn as an example, the numerical checks on the widths of contour paths $C_{\rm b}^{-}$ and $C_{\rm b}^{+}$ introduced due to the blocking effects will be discussed. As we have said, $C_{\rm b}^{-}$ and $C_{\rm b}^{+}$ should include the pole of quasiparticle energy for the blocked level and the width $E_{\rm b}$ can not be taken arbitrarily. In the following discussions, we mainly take two different blocking energy widths $E_{\rm b}$, i.e., (1) $E_{i_{\rm b}}<E_{\rm b}<-\lambda$, and (2) $E_{\rm b}>-\lambda$, where $E_{i_{\rm b}}$ is the quasi-particle energy of blocked level and $-\lambda$ is the continuum threshold. Both of them include the blocked level, but in the second case, some continuum states are included in the contour paths $C_{\rm b}^{-}$ and $C_{\rm b}^{+}$.

For the ground state of even-even nucleus $^{158}$Sn, according to the continuum Skyrme-HFB calculations,
the lowest quasiparticle state is $3p_{1/2}$ with the energy around $0.9$~MeV and the Fermi energy around $-1.0$~MeV. Thus, in the calculations for the nearby odd-even nucleus $^{159}$Sn with Green's function method, the odd neutron will be blocked on the quasiparticle state $3p_{1/2}$ and we take $E_{\rm b}=0.9$~MeV and $1.2$~MeV for discussions of the blocking energy widths.

In Fig.~\ref{Fig2}, the neutron occupation number density $n_{lj}(E)$ around the continuum threshold energy $-\lambda$ and the neutron density distributions $4\pi r^{2}\rho_{lj}(r)$ for the blocked partial wave $p_{1/2}$ of $^{159}$Sn are presented, which are calculated by the continuum Skyrme HFB with the blocking energy widths $E_{\rm b}=0.9$~MeV and $1.2$~MeV. From the occupation number density $n_{lj}(E)$ in panel (a), a discrete quasiparticle state $3p_{1/2}$ is observed around $E=0.83~$MeV below the continuum threshold $-\lambda$, and very small continuum states in the region above $-\lambda$.
Comparing the results obtained with $E_{\rm b}=0.9~$MeV and $1.2~$MeV, most of them are same except that an unphysical peak is observed in the continuum region in the case of $E_{\rm b}=1.2~$MeV, which starts from the threshold energy $E=-\lambda$ and ends at $E=1.2$~MeV. To analyse the structure of this unphysical peak, we plot in panel (b) the different contributions $n_{0,lj}(E)$, $n_{1,lj}(E)$, and $n_{2,lj}(E)$ in Eq.~(\ref{EQ:nE}) and find that it is the term $n_{2,lj}$(E) that leads to the unphysical peak in continuum. Similar problem happens also for the neutron density distributions in the coordinate space. It can be seen in panel~(c) that when the blocking energy width $E_{\rm b}=1.2$ MeV, the neutron density $4\pi r^{2}\rho_{lj}(r)$ for the partial wave $p_{1/2}$ has an increasing tail compared with that obtained with $E_{\rm b} =0.9$ MeV, which is against the outgoing decay asymptotic behavior of the nuclear wave functions. To explain the abnormal tail of density, the contributions $\rho_{0,lj}(r)$, $\rho_{1,lj}(r)$, and $\rho_{2,lj}(r)$ in Eq.~(\ref{EQ:rholj}) for $\rho_{lj}(r)$ are plotted in panel (d) and obviously it's caused by the term $\rho_{2,lj}(r)$.
However, the unphysical peak in $n_{lj}(E)$ and increasing tail in $\rho_{lj}(r)$ do not happen when taking any blocking energy width if $E_{\rm b}<-\lambda$. Thus, we can explain these problems as following: in the case of $E_{\rm b}=1.2$~MeV, extra continuum distributed over the threshold $-\lambda$ are included in blocking. For the quasiparticle states in continuum, the upper component of the wave function $\varphi_{1}(r)$ is oscillating and outgoing while the lower component $\varphi_{2}(r)$ is decaying.
Since the term $\rho_{2,lj}(r)$ in Eq.~(\ref{EQ:rholj}) is related with $\varphi^{2}_{1,n_{\rm b}l_{\rm b}j_{\rm b}}(r)$, the calculated density will be oscillating and outgoing in large coordinate if continuum states are included. Similar explanation is for the occupation number density $n_{2,lj}(E)$, which is also related with $\varphi^{2}_{1,n_{\rm b}l_{\rm b}j_{\rm b}}(r)$.

According to the above discussions, in the following, the blocking energy width $E_{\rm b}$ should be taken with $E_{i_{\rm b}} < E_{\rm b} < -\lambda$. However, note that for the very neutron rich nuclei whose Fermi surface is very close to zero, there maybe no discrete quasi-particle states and we have to block a quasi-particle state in continuum. In this case, the blocking contour path will intrude the continuum and should be taken very carefully which include only the blocked level.

\begin{figure}[t!]
 \includegraphics[width=0.4\textwidth]{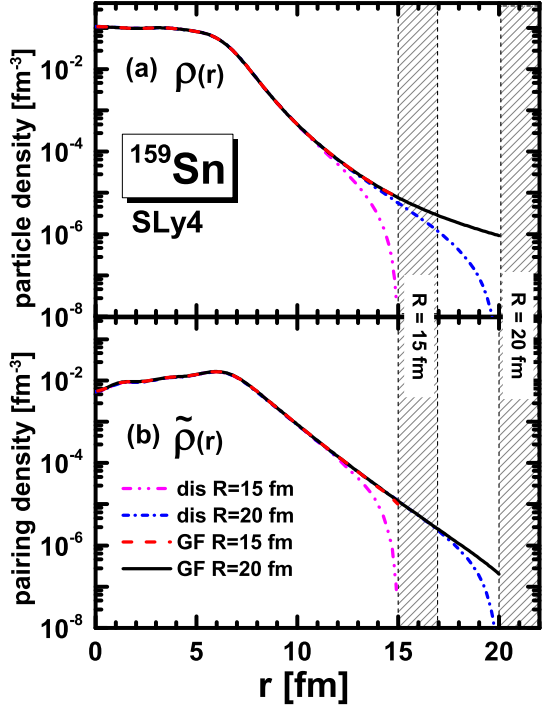}
 \caption{(a) Neutron density $\rho_{n}(r)$ and (b) neutron pairing density $\tilde{\rho}_{n}(r)$
 for $^{159}$Sn by the continuum Skyrme-HFB theory with Green's function method, in comparison with those
 by the box-discretized method. Space sizes $R=15$~fm and $R=20$~fm are taken in both calculations. The Skyrme parameter set is SLy4.}
 \label{Fig3}
\end{figure}

In the following, we will show the advantages of the Green's function method in describing the neutron-rich nuclei compared with those by box-discretized method. In Fig.~\ref{Fig3}, the particle density $\rho_{n}(r)$ and pair density $\tilde{\rho}_{n}(r)$ for neutrons in $^{159}$Sn obtained in the continuum and box-discretized Skyrme-HFB calculations are presented with coordinate space sizes of $R=15$~fm and $R=20$~fm, respectively. It can be clearly seen that in the box-discretized calculations, densities $\rho_{n}(r)$ and $\tilde{\rho}_{n}(r)$ in $^{159}$Sn decrease sharply at the edge due to the box boundary conditions which restrict wave functions being zero at the edge of the box. As a result, in order to describe the asymptotic behaviors of extended density distributions properly, large coordinate space size should be taken. However, in the continuum Skyrme-HFB calculations with Green's function method, the exponential decay of density distribution is well described. Moreover, these descriptions are independent with the space size because the correct asymptotic behaviors on the wave functions especially for the continuum states is imposed in the Green's function method to describe extended densities.

\begin{figure}[t]
 \includegraphics[width=0.45\textwidth]{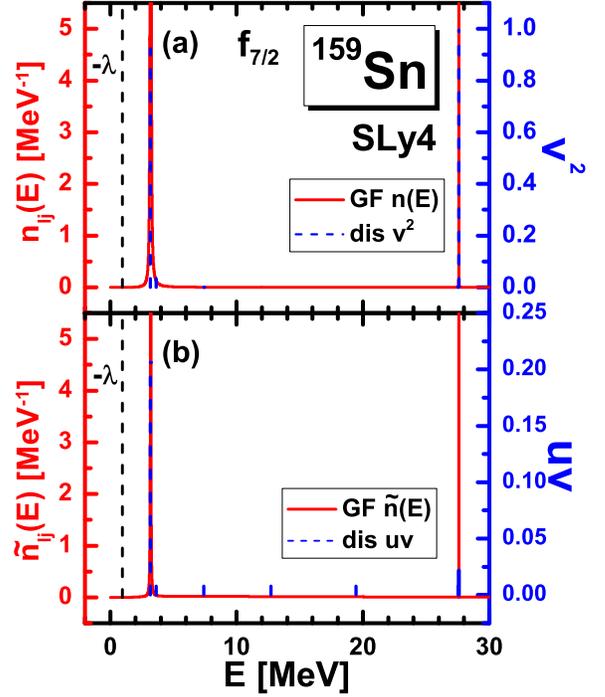}
 \caption{(a) Occupation number density $n_{lj}(E)$ and (b) pair number density $\tilde{n}_{lj}(E)$ for the partial wave $f_{7/2}$ in $^{159}$Sn obtained in the continuum Skyrme-HFB calculation with Green's function method (red solid lines). For comparison, the occupation probability $v^{2}$ and pair probability $uv$
 obtained in the box-discretzed Skyrme-HFB calculation are also plotted (blue dashed lines). The black dashed line denotes the continuum threshold $-\lambda$. The Skyrme parameter set is SLy4.}
 \label{Fig4}
\end{figure}

In Fig.~\ref{Fig4}, the occupation number density $n_{lj}(E)$ and pair number density $\tilde{n}_{lj}(E)$
obtained in the continuum Skyrme-HFB calculation with Green's function method are shown for
the partial wave $f_{7/2}$ in $^{159}$Sn displayed with red solid lines.
For comparison, the occupation probability $v^{2}\in[0,1]$ and pair probability $uv\in[0,0.25]$
in the box-discretzed Skyrme-HFB calculations are also plotted,
\begin{subequations}
\begin{eqnarray}
  v^{2}&=& \int dr~\varphi^{2}_{2,nlj}(r), \\
  uv   &=& \int dr~\varphi_{1,nlj}(r)\varphi_{2,nlj}(r),
  \label{EQ:vv}
\end{eqnarray}
\end{subequations}
where $\varphi_{1,nlj}(r)$ and $\varphi_{2,nlj}(r)$ are respectively the upper and lower components of the HFB
radial wave functions in Eq.~(\ref{EQ:rWF}).
From panel (a), two quasi-particle resonant peaks are observed in the continuum region
around quasi-particle energy $E=3.2$~MeV and $27.6$~MeV, respectively.
Especially, the state near the continuum threshold $-\lambda$ corresponds to a weakly bound single-particle
level $2f_{7/2}$ near the Fermi surface, which has a obvious width
due to the couplings with continuum while the other peak correspond to the deeply bound single-particle state $1f_{7/2}$,
the occupation number density of which is very high and sharp.
Correspondingly, the occupation probability $v^2$ denoted by blue dashed lines almost equal $1.0$ for the deeply bound $1f_{7/2}$
while less than $1.0$ for the weakly bound $2f_{7/2}$ state.
However, a series of nonphysical discrete single quasiparticle states are
also obtained with the box-discredited HFB method.
For example, the quasiparticle state $2f_{7/2}$ is discretized to three peaks by the box-discredited method.
In panel (b), the pair number density $\tilde{n}_{lj}(E)$ distribution is similar to the occupation number density $n_{lj}(E)$, except the width is obviously smaller around the Fermi energy.
In fact, it is believed that the pair number density $\tilde{n}_{lj}(E)$ represents more clearly the structure of continuum quasiparticle states due to the relevance to the pair correlation.
From the occupation number density $n_{lj}(E)$ or the pair number density $\tilde{n}_{lj}(E)$, the quasi-particle energies and widths for the quasi-particle resonant states can be read directly. The pair correlation strength, the resonant states, and the couplings between the bound states and continuum can also be investigated by analyzing the resonant widths.

\section{RESULTS AND DISCUSSION}\label{Sec:Results}

\begin{figure}[t!]
 \includegraphics[width=0.45\textwidth]{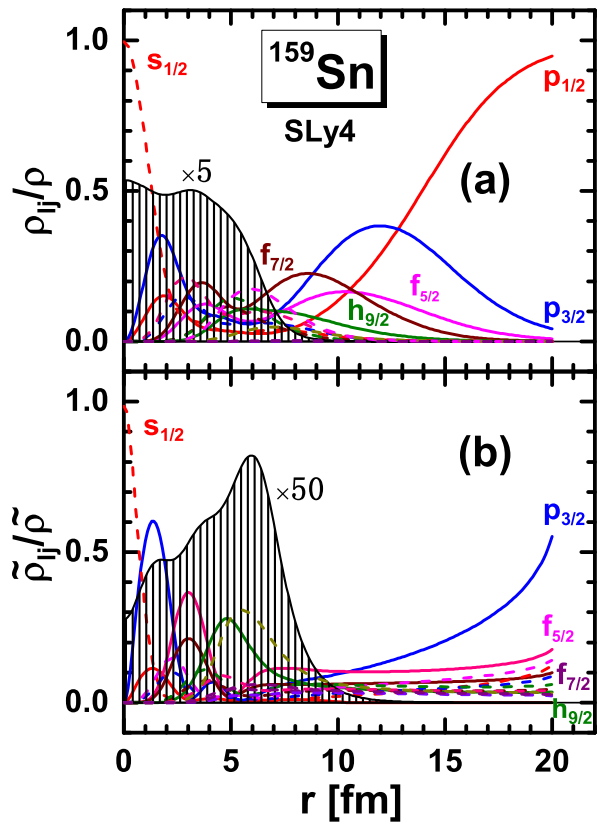}
 \caption{Compositions of different partial waves to the total partial density $\rho_{lj}(r)/\rho(r)$ (a) and the total pair density $\tilde{\rho}_{lj}(r)/\tilde{\rho}(r)$ (b) for neutrons in nucleus $^{159}$Sn by the continuum Skyrme-HFB calculations. The shallow regions corresponding to the particle density in the upper panel and the pair density in the lower panel are rescaled by multiplying a factor of $5$ and $50$, respectively. }
 \label{Fig5}
\end{figure}

\begin{figure}[t!]
 \includegraphics[width=0.45\textwidth]{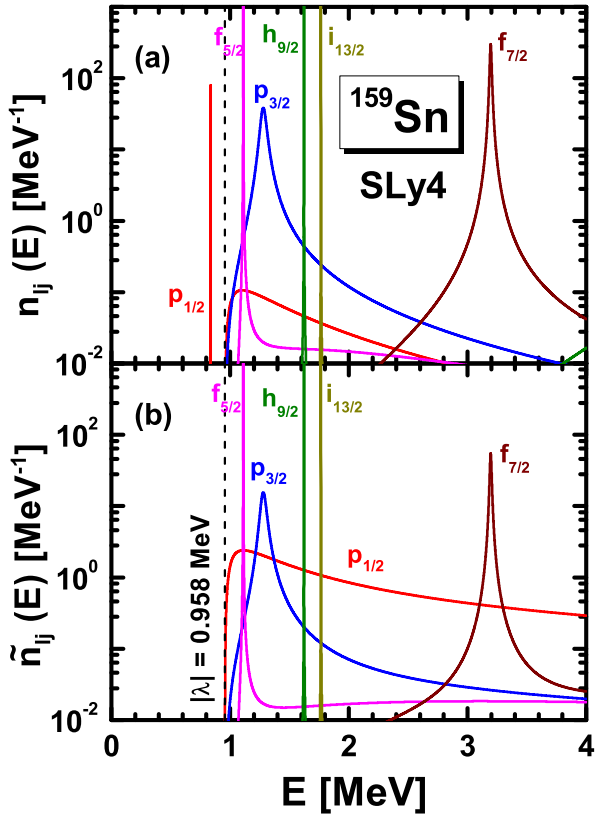}
 \caption{(a) Neutron occupation number densities $n_{lj}(E)$
and (b) neutron pair number densities $\tilde{n}_{lj}(E)$ within quasi-particle energy ranging $0$-$4$ MeV for different orbits $lj$ in $^{159}$Sn by the continuum Skyrme-HFB calculations. The black dashed line represents the threshold of the quasi-particle continuum $E=-\lambda$. }
 \label{Fig6}
\end{figure}

In this part, still taking the neutron-rich nucleus $^{159}$Sn as the example, we analyze its structure by the continuum Skyrme-HFB theory with blocking the quasi-particle state $1p_{1/2}$. In Fig.~\ref{Fig5}, the particle density $\rho(r)$ and pair density $\tilde{\rho}(r)$ for neutrons
in $^{159}$Sn as well as their contributions from different partial waves $lj$, i.e., $\rho_{lj}(r)/\rho(r)$ and $\tilde{\rho}_{lj}(r)/\tilde{\rho}(r)$, are plotted as functions of radial coordinate $r$. The shallow regions are for the total densities $\rho(r)$ and $\tilde{\rho}(r)$. The solid and dashed lines are respectively the contributions from orbits with the negative and positive parities. In panel (a), the neutron density $\rho(r)$ decreases sharply from $5$~fm and becomes very small around $9~$fm, which finally determines the neutron radius in $^{159}$Sn equal $r_{\rm n}=5.44~$fm. Besides, it can be seen clearly that
outside the nuclear surface, it's the orbits $p_{1/2}$, $p_{3/2}$, $f_{5/2}$, $f_{7/2}$, and $h_{9/2}$ that contribute a lot to the neutron density, especially the $p_{1/2}$ orbit, which is the most dominant composition in the large coordinate space with $r>15$~fm. In panel (b), the pair density $\tilde{\rho}(r)$ mainly locates around the nuclear surface due to that pairing interaction mainly effects on the orbits around the Fermi surface.

\begin{figure}[t!]
\center
 \includegraphics[width=0.45\textwidth]{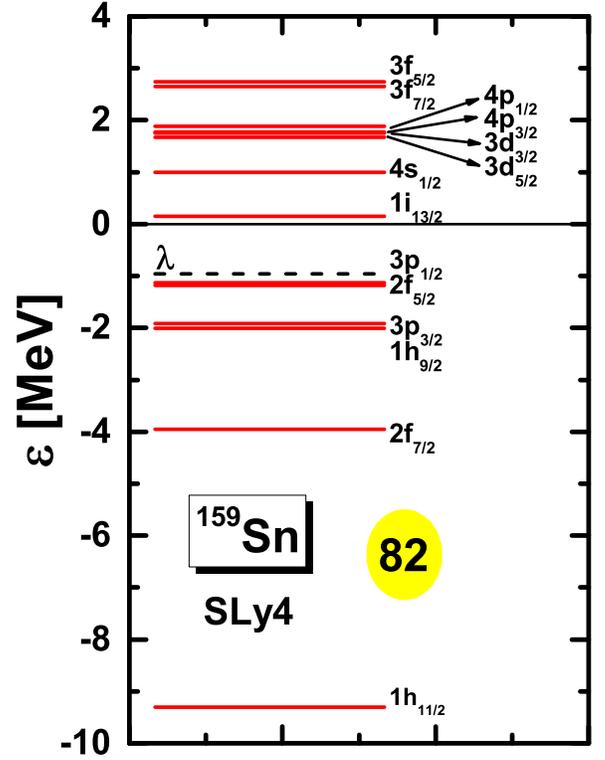}
 \caption{Neutron Hartree-Fock single-particle energy $\varepsilon$ of $^{159}$Sn around the Fermi energy. The dashed line denotes the neutron Fermi energy $\lambda$. The Skyrme parameter set is SLy4.}
 \label{Fig7}
\end{figure}

In Fig.~\ref{Fig6}, the occupation number densities $n_{lj}(E)$ and the pair number densities $\tilde{n}_{lj}(E)$ for neutrons in $^{159}$Sn are plotted in the low energy interval $E=0\sim4$ MeV. The dashed lines represent the continuum threshold with $|\lambda_{\rm n}|=0.958$~MeV, above which are quasi-particle continuum. In panel (a), one discrete quasi-particle state $1p_{1/2}$ and five resonant states $1f_{5/2}$, $1p_{3/2}$, $1h_{9/2}$, $1i_{13/2}$, and $1f_{7/2}$ are observed. All those quasi-particle states correspond to the weakly bound or continuum single-particle states around the Fermi surface as shown in Fig.~\ref{Fig7}. For the blocked quasi-particle state $1p_{1/2}$, which locates below $|\lambda_{\rm n}|$ with no width, it corresponds to the weakly bound Hartree-Fock single-particle state $3p_{1/2}$. For the quasi-particle resonant states $1f_{5/2}$, $1p_{3/2}$, $1h_{9/2}$, $1i_{13/2}$, and $1f_{7/2}$ which have peak structures with finite widths, they correspond to the single-particle states $2f_{5/2}$, $3p_{3/2}$, $1h_{9/2}$, $1i_{13/2}$, and $3f_{7/2}$, respectively.
All these single-particle states are bound except state $1i_{13/2}$. We can conclude that it is the pairing correlations that transform these bound HF single-particle orbits to quasiparticle continuum and the finite widths mainly result from the pair correlation and the couplings with continuum. In general, the pair correlation will increase the width of resonant states~\cite{PRC2012ZhangY86}. From panel (a), the quasi-particle state $1p_{3/2}$ has the largest width, which is consistent with the most important contribution for the pair density $\tilde{\rho}(r)$ at coordinate space with $r>10$~fm. The strict relations between the quasi-particle energy $E$ and single-particle energy $\varepsilon$ can be analyzed by equation $E=\sqrt{(\epsilon-\lambda)^2+\Delta^2}$ where $\Delta$ is the pairing gap. Besides, we find that it is the states around the Fermi surface that contribute the extended density distributions Fig.~\ref{Fig5}. In panel (b), the discrete state $1p_{1/2}$ disappears in terms of the pair number density while other states keep the same positions.

\section{Summary}
\label{Sec:Summary}

In this work, the self-consistent continuum Skyrme-HFB theory is extended to describe the odd-$A$ nuclei with the Green's function technique in the coordinate space. The blocking effects are incorporated by taking the equal filling approximation. Detailed formula for the densities and quasi-particle spectrum in forms of the HFB Green's function are presented for odd nuclear systems.

Taking the neutron-rich nucleus $^{159}$Sn as an example, we give the numerical details and checks. The SLy4 parameter is taken in the $ph$ channel and the DDDI is taken as the pairing interaction, the parameters of which are constrained by reproducing the experimental neutron pairing gaps for the Sn isotopes and the scattering length $a=-18.5$~fm in the $^1S$ channel of the bare nuclear force. To perform the integrals of the Green's function, three contour paths $C_{E<0}$, $C_{\rm b}^{-}$, and $C_{\rm b}^{+}$ are chosen, the height of which are taken uniformly $\gamma=0.1$~MeV, and width of $C_{E<0}$ is taken as the maximal quasi-particle energy $E_{\rm cut}=60$~MeV to enclose all the negative quasiparticle energies. The numerical checks on the widths of contour paths $C_{\rm b}^{-}$ and $C_{\rm b}^{+}$ introduced for the blocking effects are discussed and it is found that the width $E_{\rm b}$ should taken with $E_{i_{\rm b}}<E_{\rm b}<-\lambda$. This means that the contour paths $C_{\rm b}^{-}$ and $C_{\rm b}^{+}$ should include the blocked quasi-particle state but can not intrude to the continuum area.  Besides, by comparing with the box-discretized Skyrme-HFB calculations, the advantages of the Green's function method in describing the neutron-rich nuclei are shown. First, Green's function method can describe the extended density distributions very well and these descriptions are independent with the space size. Second, Green's function method can describe the quasiparticle spectrum especially the continuum very well, by which the energies and widths of quasi-particle resonant states can be given directly.

Finally, we investigated the halo structure of the neutron-rich nucleus $^{159}$Sn with the continuum Skyrme-HFB theory by blocking the quasi-particle state $1p_{1/2}$. We find that it is the weakly bound states $3p_{1/2}$, $2f_{5/2}$, $3p_{3/2}$, $1h_{9/2}$, and $2f_{7/2}$ that contribute a lot for the extended density distributions at large coordinate space. Besides, the particle number density $n_{lj}(E)$ and pair number density $\tilde{n}_{lj}(E)$ are also studied, from which the quasi-particle energies and the width of resonant states can be extracted. The pairing correlation
and the couplings with the continuum can be analyzed from the width of quasi-particle resonant states.

\begin{acknowledgments}
T.-T. S. is grateful to Prof. J. Meng, Prof. M. Matsuo, and Dr. Y. Zhang for
fruitful discussions. This work was partly supported by the National Natural Science Foundation of China (Grant No.~11505157 and No.~11705165) and the Physics Research and Development Program of Zhengzhou University (Grant No.~32410017).
\end{acknowledgments}


\end{document}